# Designing a Holistic At-Home Learning Aid for Autism

**Catalin Voss**  catalin@cs.stanford.edu
**Nick Haber**  nhaber@stanford.edu
**Peter Washington**  peterwashington@stanford.edu
**Aaron Kline**  akline@stanford.edu
**Beth McCarthy**  bethmac@stanford.edu
**Jena Daniels**  danielsj@stanford.edu
**Azar Fazel**  azarf@stanford.edu
**Titas De**  titasde@stanford.edu
**Carl Feinstein**  carlf@stanford.edu
**Terry Winograd**  winograd@cs.stanford.edu
**Dennis Wall**  dpwall@stanford.edu

Stanford University
Stanford, CA 94305, USA

## Abstract
In recent years, much focus has been put on employing technology to make novel behavioral aids for those with autism. Most of these are digital adaptations of tools used in standard behavioral therapy to enforce normative skills. These digital counterparts are often used outside of both the larger therapeutic context and the real world, in which the learned skills might apply. To address this, we are designing a system of automatic expression recognition on wearable devices that integrates directly into the families' daily social interactions, to give children and their caregivers the tools and information they need to design their own holistic therapy. In order to develop a tool that will be truly useful to families, we proactively include children with autism and their families as co-designers in the development process. By providing an app and interface with interchangeable social feedback options, we aim to produce a framework for therapy that folds into their daily lives, tailored to their specific needs.

## ACM Classification Keywords
H.5.m. Information interfaces and presentation (e.g., HCI): Miscellaneous

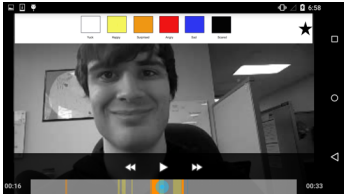

**Parent Review System**

**Figure 1**: Users and their caregivers have the ability to review pre-curated videos of their at-home "sessions" of emotional interaction. While reviewing the videos, users have the ability to reflect on specific behavior and mark particular parts of the session as important or write comments.

**Introduction**

Autism is quickly becoming a public health crisis, with 1 in 68 children affected by Autism Spectrum Disorder (ASD) today [2]. Many of these children experience impairments in the acquisition of social skills ranging from difficulty with recognizing facial expressions and making eye contact to struggles with interpreting interpersonal interactions [5]. This is true even for young adults with very mild autistic traits [7]. The most effective solution available today for treatment of autism symptoms, including social functioning, is intensive behavioral intervention [1]. To-date, efforts to develop assistive technological aids for autism therapy mostly translate existing behavioral intervention tools, such as flashcard therapy, a common intervention involving painstaking memorization of facial expressions, into the digital domain (see for example [3]).

These approaches to delivery of traditional behavioral intervention target specific cognitive challenges (e.g. children's ability to recognize emotions) and aim to increase a child's social awareness, often integrated into the context of a broader therapy. While they have some documented therapeutic success [1], they are deployed in artificial environments, removed from the contexts in which the social skills taught must actually be used. As a consequence, they have at least two major drawbacks:

1. Suffering from a reductionist perspective of attempting to cure various perceived deficits outside of the context in which the difficulty manifests, these techniques may change behaviors through artificial training that is irrelevant to the overall improvement of quality of life.
2. Even in cases in which addressing the specific perceived deficit is warranted, the separation from natural social contexts makes the transfer of new skills to the real world more difficult.

Further, when flashcard-type interventions do work, they are often successful as part of a more holistic behavioral therapy that considers the child's individual family circumstances. Digitized forms of interventions, when distributed widely and removed from the context of therapy, easily lose their value.

The *Autism Glass Project* aims to address these difficulties by taking a more holistic approach. We built a learning aid, co-designed by people with autism, that integrates into the daily lives of families. Following previous work to design just-in-time in-situ learning aids for facial affect by presenting emotional cues in a fun and intuitive way on a mini-computer [6], we have proposed a system for automatic facial expression recognition that runs on wearable glasses and delivers real time social cues to individuals with ASD in their natural environment. Our system includes various gamified activities that children can use during informal behavioral therapy sessions and provides feedback to caregivers and children about their progress in the form of recorded and pre-curated videos of actual interaction, as well as metrics like eye contact insights through an Android app. Rather than enforcing normative behavior and directly suggesting specific actions as the result of our emotion recognition aid, we aim to give users more socially-relevant information in real time. By providing a detailed behavioral review (Fig. 1), we then enable users to re-examine the social interaction along with the real-time feedback they received. This review unlocks the child's social awareness and design-space for building more informed responses to the social situations they encountered. The details of our system are described in [9]. Through a multi-stage study in which we design the system iteratively with small groups of participants before scaling to a larger cohort for validation, we believe that we can create a program that will go beyond the targeting of any particular learning deficit and will be broadly useful to families with autism.

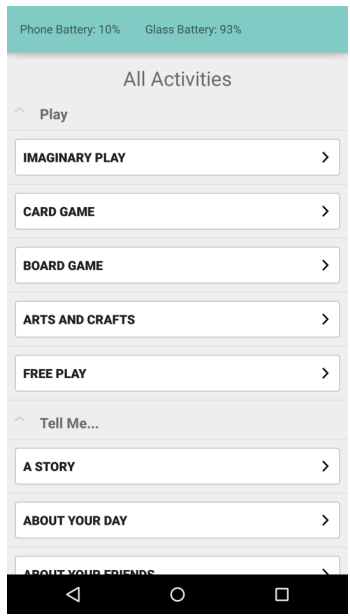

**Figure 2**: "Example activities" interface that is presented when users start an "unstructured activity" session in the Android application used to control the Autism Glass system.

## Designing an Interactive Learning Aid

By providing cues for identifying emotions through an unobtrusive, mobile device in real time, the Autism Glass system can be incorporated into a dynamic, holistic approach to improving social skills. Children wearing the glasses are able to learn and practice emotion recognition in naturally-occurring contexts. We hypothesize that this will facilitate understanding of the cause and effect of emotions and their associated mental states more effectively than static learning methods that are divorced from the context in which emotions actually occur. To turn our prototype into a successful learning aid, we have set up our system and a design study with 90 participants with autism according to the following paradigms:

- The social aid should capture data about both people interacting with the child and his or her reaction to them.
- We include "example activities" through the Android app (see Fig. 2) and we do not limit families' use of the system beyond encouraging them to use it over short periods of time during social interaction.
- The system offers an artificial intelligence service, namely facial expression recognition, meant to inform behavioral patterns, and in the process, captures data that we can later review and use for weakly-labeled training for relevant, perhaps broader, tasks of social action recognition. Should families alert us to relevant behaviors that we are not currently tracking, this dataset provides a starting point for developing classifiers for these as well.
- We give families full control over the review of their behavioral data in an accessible, pre-curated way and provide tools to organize this data, e.g. by allowing participants to write comments about their "sessions." For the task of emotion recognition, our video review app "color-codes" the video scrub bar with the emotions recognized (Fig. 1).

We believe this setup will allow subjects to openly explore the possibilities of this technology for therapy. As we work closely with participants ("co-designers") of this system, we hope to learn the ways in which the system can best serve families' specific needs. Video data captured in this fashion will enable us to pay close attention to the nature of device use (frequency, context, etc.), issues encountered, and areas for improvement. As participants use the device at home, they communicate with study coordinators on a regular basis via an in-app messaging system and participate in bi-weekly semi-structured interviews in the lab. This feedback, coupled with the video data and the various personalizing choices participants make, allows us to explore and address the positive and negative aspects of our system in order to create a successful, systematized at-home therapy. As we move from design, to validation, to deployment of the system as a treatment, data capture and personalization will allow us to tailor the experiences to participants' needs.

Following in-lab interaction with our device [4], some participants and their caregivers reported that they used our aid to verify their interpretation of what another person is emoting. This allowed them to become more comfortable engaging in conversations. If they can confirm the person in front of them is indeed experiencing a certain emotion, then they are more empowered to respond in an informed manner. Other participants reported that the most useful aspect of our tool was that it alerted them to the face in the first place. One caregiver of an older participant found the emotion recognition aspect of our tool "perhaps useful a few years ago" and believed that the "behavioral review" would be the most useful intervention for them now. Parents of participants of different developmental ages saw different components of our solution as most beneficial.

## Conclusion

We propose to use an egocentric wearable to capture video data on actual behavioral interaction and offer an artificial intelligence service meant to inform behavioral patterns. We believe that this paradigm allows caregivers to provide more holistic therapy to children with autism. As we think about building learning aids as designers, engineers, and clinicians, we consider it essential to take a more holistic approach, even when providing therapy targeted to a very specific behavioral deficit. Without built-in feedback mechanisms that put much of the therapy into the control of caregivers (e.g. using the video and behavioral data review), we do not believe that the Autism Glass aid would be useful outside of the context of structured behavioral therapy.

As the gap between the number of affected individuals with autism and the number of available behavioral therapists expands, we must build in comprehensive feedback interfaces when building technological learning aids, as we cannot conceivably rely on therapists to provide that holistic context.

Finally, we believe that the iterative design-before-validation approach we have taken in our study – video data capture, regular interviews, and device personalization – is a key ingredient in enabling us to build a useful tool that can be integrated into the daily lives of families.

By framing one of the symptoms of Autism as a deficit in social awareness, we, as designers, can aim to provide affected people with socially relevant information in real time, and give them and their caregivers the opportunity to review and inform their behavioral decisions. This paradigm enables families to develop their own therapy rather than settle for normative reinforcement, for which many aids are available.

## Acknowledgements

This work was supported with funding from the Hartwell Foundation, Dekeyser and Friends Foundation, and by the David and Lucile Packard Foundation Special Projects Grant under grant 2015-62349. We would also like to thank Google for donating 35 units of Google Glass through the Google Gifts Material Grant.

## References

1. Dawson, G. and Burner, K., 2011. Behavioral interventions in children and adolescents with autism spectrum disorder: a review of recent findings. Current opinion in pediatrics, 23(6), pp.616-620.

2. Developmental, D.M.N.S.Y. and 2010 Principal Investigators, 2014. Prevalence of autism spectrum disorder among children aged 8 years-autism and developmental disabilities monitoring network, 11 sites, United States, 2010. Morbidity and mortality weekly report. Surveillance summaries (Washington, DC: 2002), 63(2), p.1.

3. Golan, O., Ashwin, E., Granader, Y., McClintock, S., Day, K., Leggett, V. and Baron-Cohen, S., 2010. Enhancing emotion recognition in children with autism spectrum conditions: An intervention using animated vehicles with real emotional faces. Journal of autism and developmental disorders, 40(3), pp.269-279.

4. Haber, N., Voss, C., Fazel, A., Daniels, J., Tanaka, S., Winograd, T., Feinstein, C., Wall, D., 2016. Autism Glass: Wearable Behavioral Aid Augments Social Learning in Children with Autism. In submission.

5. Hobson, R.P., 1995. Autism and the development of mind. Psychology Press.

6. Madsen, M., El Kaliouby, R., Goodwin, M. and Picard, R., 2008, October. Technology for just-in-time in-situ learning of facial affect for persons diagnosed with an autism spectrum disorder. In Proceedings of the 10th international ACM SIGACCESS conference on Computers and accessibility (pp. 19-26). ACM.

7. Poljac, E., Poljac, E. and Wagemans, J., 2013. Reduced accuracy and sensitivity in the perception of emotional facial expressions in individuals with high autism spectrum traits. Autism, 17(6), pp.668-680.

8. Ruble, L.A., Heflinger, C.A., Renfrew, J.W. and Saunders, R.C., 2005. Access and service use by children with autism spectrum disorders in Medicaid managed care. Journal of Autism and Developmental Disorders, 35(1), pp.3-13.

9. Washington, P., Voss, C., Haber, N., Tanaka, S., Daniels, J., Feinstein, C., Winograd, T. and Wall, D.P., 2016. A Wearable Social Interaction Aid for Children with Autism. To appear in ACM annual conference on Human Factors in Computing Systems (CHI) Late-Breaking Work.